\documentclass[paper,showpacs,preprintnumbers,amsmath,amssymb]{revtex4}

\usepackage{graphicx}
\usepackage{dcolumn}
\usepackage{bm}

\newcommand{\be}{\begin{equation}}
\newcommand{\en}{\end{equation}}
\newcommand{\bea}{\begin{eqnarray}}
\newcommand{\ena}{\end{eqnarray}}

\begin{document}


\title{ Extended Curvaton reheating in inflationary models }

\author{Cuauhtemoc Campuzano}
 \email{ccampuz@mail.ucv.cl}
\affiliation{ Instituto de F\'{\i}sica, Pontificia Universidad
Cat\'{o}lica de Valpara\'{\i}so, Casilla 4059, Valpara\'{\i}so,
Chile.}
\author{Sergio del Campo}
 \email{sdelcamp@ucv.cl}
\affiliation{ Instituto de F\'{\i}sica, Pontificia Universidad
Cat\'{o}lica de Valpara\'{\i}so, Casilla 4059, Valpara\'{\i}so,
Chile.}
\author{Ram\'on Herrera}
\email{ramon.herrera.a@mail.ucv.cl} \affiliation{ Instituto de
F\'{\i}sica, Pontificia Universidad Cat\'{o}lica de
Valpara\'{\i}so, Casilla 4059, Valpara\'{\i}so, Chile.}

\date{\today}

\begin{abstract}
 The curvaton reheating in a non-oscillatory  inflationary
universe model  is studied in a Jordan-Brans-Dicke theory. For
different scenarios, the temperature of reheating is computed. The
result tells us that the reheating temperature becomes practically
independent of the Jordan-Brans-Dicke parameter $w$. This
reheating temperature results to be quite different when compared
with that obtained from Einstein`s theory of gravity.
\end{abstract}

\maketitle

\section{Introduction}

It is well known that many long-standing problems of the Big Bang
model (horizon, flatness, monopoles, etc.) may find a natural
solution in the framework of the inflationary universe model
\cite{guth,infla}.

One of the successes of the inflationary universe model is that it
provides a causal interpretation of the origin of the observed
anisotropy of the cosmic microwave background (CMB) radiation, and
also the distribution of large scale structures \cite{astros}. In
standard inflationary universe models, the acceleration of the
expansion of the universe is driven by a scalar field $\psi$
(inflaton) with a specific scalar potential, and the quantum
fluctuations associated with this field generate the density
perturbations seeding the structure formation at late time in the
evolution of the universe.  To date, the accumulating
observational data, especially those coming from the CMB
observations of the WMAP satellite \cite{astros}, indicate that
the power spectrum of the primordial density perturbations becomes
nearly scale-invariant, just as predicted by the single-field
inflationary model .

At the end of inflation the universe is typically in a highly
non-thermal state. An exception is the warm inflation scenario,
where there is particle production  during inflation
\cite{berera}. The key ability of inflation is to homogenize the
universe, which means that it leaves the universe at very low
temperature and hence any successful theory of inflation must also
explain how the universe was reheated - or perhaps defrosted - to
the Big Bang picture \cite{Lyth1}. This approach must include
baryogenesis and nucleosynthesis, the latter implying that the
temperature must have been higher than 1 MeV and the former
requiring energies significantly higher.

One path to defrost the universe after inflation is known as
reheating \cite{Kolb1}. Elementary theory of reheating was
developed in \cite{Dolgov} for the new inflationary scenario.
During reheating, most of the matter and radiation of the universe
are created, usually via the decay of the scalar field that drives
inflation, while the temperature grows in many orders of
magnitude. Of particular interest is a quantity known as the
reheating temperature. The reheating temperature  is associated
with the temperature of the universe when the Big Bang scenario
begins, that is, when the radiation epoch begins. In general, this
epoch is generated by the decay of the inflaton field, which leads
to a creation of particles of different kinds.

The stage of oscillations of the scalar field is an essential part
of the standard mechanism of reheating. However, there are some
models where the inflaton potential does not have a minimum and
the scalar field does not oscillate. Here, the standard mechanism
of reheating does not work \cite{Kofman}. These models are known
in the literature as non-oscillating models, or simply NO models
\cite{fengli,Felder1}. The NO models correspond to runaway fields
such as {\bf moduli} fields in string theory which are potentially
useful for inflation model-building because they present flat
directions which survive the famous $\eta$-problem of inflation
\cite{dine}. On the other hand, an important use of NO models is
quintessential inflation, in which the tail of the potential can
be responsible for the accelerated expansion of the present
universe \cite{Dimopoulus}.

However, these models present another type of $\eta$-problem,
which has to do with the fact that between the inflationary
plateau and the quintessential tail there is a difference of over
a hundred orders of magnitude. According to the above, it is
useful to consider an exponential potential\cite{Dimopoulus}.
During inflation, the power law expansion may be realized  if the
inflaton field  with an exponential potential dominates the energy
density of the universe. Originally, this model was described in
terms of Einstein`s theory of gravity in Ref.\cite{Lucchin}.
However, during the past decade, a great number of studies based
on a less standard theory have been carried out, namely, the
Jordan-Brans-Dicke (JBD) theory \cite{Jbd}.

JBD theory is characterized by the presence of a dynamic massless
scalar field (the JBD field) which couples directly with the
metric in the gravitational sector, providing a variable
gravitational constant. The theory explicitly presents a
nonminimal coupling between the scalar JBD field and the scalar
curvature. In Ref.\cite{Susperregi} is given a detailed discussion
about inflation in the context of a JBD theory, for an exponential
potential.

The first mechanism of reheating  for NO models in general
relativity theory was the gravitational particle production
\cite{ford}, but this mechanism is quite inefficient, since it may
lead to certain cosmological problems \cite{ureña,Sami_taq}. An
alternative mechanism of reheating in NO models is the instant
preheating, which introduces an interaction between the scalar
field responsible for inflation with another scalar field
\cite{Felder1}. Another possibility for reheating in NO models is
the introduction of the curvaton field, $\sigma$ \cite{ref1u},
which has recently received a lot of attention in the literature
\cite{L1,L2}. The curvaton approach is an interesting new proposal
for explaining the observed large-scale adiabatic density
perturbations in the context of inflation. Here, the hypothesis is
such that the adiabatic density perturbation is originated  from
the  ``curvaton field" and not from the inflaton field. In this
scenario, the adiabatic density perturbation  is generated only
after inflation, from an initial condition which corresponds to a
purely isocurvature perturbation \cite{Mollerach}.

On the other hand, the decay of the curvaton field into
conventional matter offers an efficient mechanism of reheating.
The curvaton field has the property that its energy density is not
diluted during inflation, so that the curvaton may be responsible
for some or all of the matter content of the universe at present.

In this paper we shall explore an application of curvaton
reheating in a JBD theory for a NO model. Specifically we explore
the model to an exponential scalar potential.

We follow a similar procedure described in
Refs.\cite{ureña,cdh1,cdh2}. As the energy density decreases, the
inflaton field makes a transition into a kinetic energy dominated
regime bringing inflation to an end.  We consider the evolution of
the curvaton field through three different stages. Firstly, there
is a period in which the inflaton energy density is the dominant
component, i.e, $\rho_\psi\gg\rho_\sigma$, even though the
curvaton field survives the rapid expansion of the universe. The
following stage i.e., during the kinetic epoch \cite{refere3}, is
that in which the curvaton mass becomes important. In order to
prevent a period of curvaton-driven inflation, the universe must
remain inflaton-driven until this time. When the effective mass of
the curvaton becomes important, the curvaton field starts to
oscillate around the minimum of its potential. The energy density,
associated with the curvaton field, starts to evolve as
non-relativistic matter.

At the final stage, the curvaton field decays into radiation and
then the standard Big Bang cosmology is recovered afterwards. In
general, the decay of the curvaton field should occur before
nucleosynthesis happens. Other constraints may arise depending on
the epoch of the decay, which is governed by the decay parameter,
$\Gamma_\sigma$. There are two scenarios to be considered,
depending on whether the curvaton field decays before or after it
becomes the dominant component of the universe.

In section \ref{secti} the inflationary dynamics in a JBD theory
is described. In section \ref{sectii} the dynamics of the extended
model in the kinetic epoch is developed. Section \ref{sectiii}
studies the dynamics of the curvaton field through different
stages. In section \ref{sectgw} some constraints from
gravitational waves in the kinetic epoch are described. Finally we
present the conclusions section \ref{conclu}.

\section{Extended Inflationary Model \label{secti}}

The dynamics of the Friedman-Robertson-Walker cosmology in the JBD
theory, is described by the equations
\begin{equation}
\ddot{\Phi}+3H\dot{\Phi}=\frac{8\pi}{(2w+3)}(\rho_\psi-3
P_\psi)\;, \label{key_01}
\end{equation}

\begin{equation}
 H^2\,+H\frac{\dot{\Phi}}{\Phi}=\frac{8\pi}{3}\frac{1}{\Phi}\rho_\psi+
 \frac{w}{6}\left(\frac{\dot{\Phi}}{\Phi}\right)^2\label{key_02},
\end{equation}
and \begin{equation} \ddot{\psi}+3H\dot{\psi}=-\frac{\partial
V(\psi)}{\partial\psi},\label{3}
\end{equation}
where $\Phi$ denotes the JBD scalar field  (with unit $m_p^2$,
where $m_p$ is the Planck mass), $H=\dot{a}/a$ is the Hubble
factor, $a$ is a scale factor, $\psi$ the standard inflaton field,
$V(\psi)$ is the effective potential associated with this field,
and we assume to be
\begin{equation*}
V(\psi)=V_0 e^{-\alpha\psi},
\end{equation*}
where $\alpha$ and $V_0$ are free parameters. In the following we
shall take $\alpha> 0$ (with unit $1/m_p$). In Einstein`s theory
of gravity, power law inflation may take place if
$\alpha<\sqrt{2}/m_p$\cite{Lucchin}. Here, $P_\psi$ represents the
pressure associated with the infaton field, and $w=Constant$
corresponds to the JBD parameter. As is mentioned in
Ref.\cite{omega}, experimental tests of weak-field in the
solar-system have constrained the post-Newtonian deviation from
Einstein gravity, where it was found that the JBD parameter should
satisfy the inequality $w>500$. According to recent reports, this
bound would increase to several thousands \cite{omega121314} .
Moreover, the Einstein`s theory of gravity is recovered in the
limits $w\longrightarrow\infty$ and $\Phi=Cte.=m_p^2$. Dots mean
derivatives with respect to time and we use  units in which
$c=\hbar=1$.

During inflation the $\ddot{\Phi}$ term in Eq.(\ref{key_01}) can
be ignored in the sense that $\ddot{\Phi}\ll H\dot{\Phi}$. Also,
$\dot{\Phi}/\Phi\ll H$ (because $w\gg 1$) and $\rho_\psi\sim
V(\psi)$, since $\dot{\psi}^2/2 < V(\psi)$. Under this
approximation the field equations (\ref{key_01}) and
(\ref{key_02}) become,
\begin{eqnarray}
3H\dot{\Phi}&=&\frac{32\pi}{2w+3}V, \label{inf1}\\
H^2&=&\frac{8\pi}{3}\frac{V}{\Phi}. \label{inf2}
\end{eqnarray}
Now, from Eq.(\ref{3}) we get, under the slow roll approximation
for the inflaton field $\psi$ ($\ddot{\psi}$ is negligible)

\begin{equation}
3H\dot{\psi}=-\frac{\partial V}{\partial \psi}, \label{inf3}
\end{equation}
and combining the Eqs. (\ref{inf1}) and (\ref{inf2}) we have,
\begin{equation}
\frac{\dot{\Phi}}{\Phi}=\frac{4 H}{2w+3}, \label{FI}
\end{equation}
which gives
\begin{equation}
\frac{\Phi}{\Phi_i}=\left(\frac{a}{a_i}\right)^{\beta},\label{eq8}
\end{equation}
where $\beta=4/(2w+3)$. We see that the JBD field $\Phi$ increases
during the inflationary epoch. Now, from Eqs. (\ref{inf2}),
(\ref{inf3}) and (\ref{eq8}), we get
\begin{equation}
\frac{\psi}{\psi_i}=\left(\frac{a}{a_i}\right)^\beta=\frac{\Phi}{\Phi_i},
\end{equation}
where we have chosen the initial condition
$\Phi_i=8\pi\psi_i\beta/\alpha$.

During  the  inflationary epoch we could write
$H_i^2=H_f^2\frac{\Phi_f V_i}{\Phi_i
V_f}=H_f^2\frac{\Phi_f}{\Phi_i}e^{-\alpha(\psi_i-\psi_f)}$, where
the subscripts $i$ and $f$ are used to denote the beginning and
the end of inflation, respectively. Inflation ends when the slow
roll  condition, is not satisfied anymore, i.e.
\begin{equation*}
\epsilon_f=-\frac{\dot{H}}{H^2}=\left[\frac{2}{(3+2w)}+\frac{\Phi}{16\pi}
\left(\frac{V_{,\psi}}{V}\right)^2\right]_f\simeq 1,
\end{equation*}
from which we get that
$\Phi_f=\frac{16\pi}{\alpha^2}(1-\beta/2)=\frac{8\pi\beta}{\alpha}\psi_f>0$,
i.e. $\beta$ is less than 2. There, we have denoted
$V,_\psi=\partial V/\partial\psi $. The initial value of the
square Hubble factor as a function of the total number of
e-foldings, $N_0=\ln(a_f/a_i)$, becomes

\begin{equation} H_i^2=\frac{\alpha^2 V_0}{3(2-\beta)}
e^{(N_0\beta-\alpha\psi_i)}.\label{hi}
\end{equation}

\section{Kinetic Epoch\label{sectii}}

When inflation has finished, the term $\partial V/\partial\psi$ is
negligible compared to the friction term. This epoch is called the
`kinetic epoch' or `kination' \cite{refere3}, and we will use the
subscript `k' to label the values of the different quantities at
the beginning of this epoch. The kinetic epoch does not occur
immediately after inflation; there may exist a middle epoch where
the potential force is negligible with respect to the friction
term \cite{Guo}. In the kinetic epoch we have
$\dot{\psi}^2/2>V(\psi)$ corresponding to the relation
$P_\psi=\rho_\psi$ which represents a stiff fluid.

The dynamics of the Friedman-Robertson-Walker cosmology in the JBD
theory, in the kinetic epoch, is described by the equations
\begin{equation}
\ddot{\Phi}+3H\dot{\Phi}=-\frac{8\pi}{(2w+3)}\;\dot{\psi}^2,
\label{ecphi}
\end{equation}

\begin{equation}
 H^2\,+H\frac{\dot{\Phi}}{\Phi}=\frac{8\pi}{3}\frac{1}{\Phi}\frac{\dot{\psi}^2}{2}+
 \frac{w}{6}\left(\frac{\dot{\Phi}}{\Phi}\right)^2\label{key_2},
\end{equation}
 and
\begin{equation}
\ddot{\psi}+3H\dot{\psi}=0\Longrightarrow\;\dot{\psi}=\frac{\dot{\psi}_k
a_k^3}{a^3}\label{solpsi}.
\end{equation}

The energy density of the inflaton field is defined by
$\rho_{\psi}=\dot{\psi}^2/2+V(\psi)\simeq \dot{\psi}^2/2$ and,
since it has a behavior like stiff matter, we get that
$\rho_{\psi}\propto(1/a)^6 $.

 From Eqs.(\ref{ecphi}) and (\ref{solpsi}), we have

\begin{equation}
\ddot{\Phi}+3H\dot{\Phi}=-\frac{8\pi}{(2w+3)}\;\frac{\dot{\psi}_k^2
a_k^6}{a^6}. \label{key_1}
\end{equation}
If we introduce a new variable, $dt=a^3 d\eta$, Eq.(\ref{key_1})
is solved and gives \cite{Barrow}
\begin{equation}
\Phi=\Phi_k-\frac{4\pi}{(2w+3)}\dot{\psi_k}^2 a_k^6
(\eta-\eta_k)^2,\label{p}
\end{equation}
where we have chosen  the initial conditions
$\left.\frac{d\Phi}{d\eta}\right|_{\eta=\eta_k}=\left.\Phi'
\right|_{\eta=\eta_k}=\Phi'_k=0$. We note that, during the kinetic
epoch , $\Phi'<0$ as it could be seen from Eq.(\ref{p}). In this
period the JBD field, $\Phi(\eta)$, is greater than the value
$\Phi_0=\frac{1}{G}=m_p^2$. In this way the JBD field lies in the
range $\Phi_k>\Phi(\eta)>\Phi_0$, where $\Phi_0$ is the actual
value of the JBD field and $G$ is the Newton constant.

Using the solutions given by Eqs. (\ref{solpsi}) and (\ref{p}), we
may write
\begin{equation}
\rho_{\psi}=\rho_{\psi}^{k}\frac{a_{k}^6}{a^6} \label{rho},
\end{equation}
and
\begin{equation}
H(a,\Phi)=H=H_{k}\frac{a_k^3}{a^3}\left[\frac{\sqrt{\Phi_k}}
{\sqrt{1+2w/3}}\left(\frac{\sqrt{(\Phi_k-\Phi)}}{\Phi}+
 \sqrt{\Phi_k\left[1+\frac{2w}{3}\right]}\;\;\frac{1}{\Phi}\right)
 \right]
  \label{h}.
\end{equation}
Note that, in the limit $w\longrightarrow\infty$ and
$\Phi=Cte.=\Phi_k=m_p^2$, we obtained
  the Hubble factor in the GR theory which  follows the law
$H^{(GR)}\propto(1/a)^3$.

\section{Curvaton Field \label{sectiii}}

We now study the dynamics of the curvaton field, $\sigma$, through
different stages. This study allows us to find some constraints on
the parameters and thus, to have a viable curvaton scenario. We
considered that the curvaton field obeys the Klein-Gordon equation
and, for simplicity, we assume that its scalar potential
associated with this field is given by
\begin{equation}
U(\sigma)=\frac{m^2\sigma^2}{2}\;,
\end{equation}
where $m$ is the curvaton mass.

First of all, it is assumed that the  energy density
$\rho_{\psi}$, associated with the inflaton field, is the dominant
component when it is compared with the curvaton energy density,
$\rho_\sigma$. In the next stage, the curvaton field oscillates
around the minimum of the effective potential $U(\sigma)$. Its
energy density evolves as a non-relativistic matter and, during
the kinetic epoch, the universe remains inflaton-dominated. The
latter stage corresponds to the decay of the curvaton field into
radiation and then the standard Big-Bang cosmology is recovered.

In the inflationary regime it is supposed that the  curvaton field
is effectively massless and its dynamics is described in detail in
Refs.\cite{dimo,postma,ureña}. During inflation, the curvaton
would roll down its potential until its kinetic energy is depleted
by the exponential expansion and only then, i.e. only after its
kinetic energy has almost vanished, it becomes frozen and assumes
roughly a constant value, i.e. $\sigma_*\approx \sigma_f$. The
subscript $``*''$ here refers to the epoch when the cosmological
scales exit the horizon.

The hypothesis is that during the kinetic epoch  the Hubble
parameter decreases so that its value is comparable with the
curvaton mass, i.e. $m\simeq H$ (the curvaton field becomes
effectively massive). From Eq.(\ref{h}), we obtain

\begin{equation}
\frac{m}{H_{k}}=\frac{a_k^3}{a_m^3}\left[\frac{\sqrt{\Phi_k}}
{\sqrt{1+2w/3}}\;\;\frac{1}{\Phi_m}\left(\sqrt{(\Phi_k-\Phi_m)}+
 \sqrt{\Phi_k\left[1+\frac{2w}{3}\right]}\;\;\right)\right],\label{mh}
\end{equation}
where the `m' label represents the quantities at the time when
the curvaton mass is of the order of $H$ during the kinetic epoch.

In order to prevent a period of curvaton-driven inflation, the
universe must still be dominated by the inflaton matter, i.e.
$\rho_{\psi}|_{a_m}=\rho_{\psi}^{(m)}\gg\rho_{\sigma}(
\sim\,U(\sigma_f)\simeq\,U(\sigma_*))$ .  This inequality allows
us to find a constraint on the values of the curvaton field
$\sigma_*$ in the epoch when the cosmological scales exit the
horizon. Then, from Eq.(\ref{key_2}) at the moment when $H\simeq
m$, we obtain the inequality

\begin{equation}
\frac{m^2\sigma_*^2}{2\rho_\psi^{(m)}}\ll 1 \Rightarrow
\sigma_*^2\ll\frac{3\Phi_m}{4\pi}\left[1+\frac{1}{m}
\left(\frac{\dot{\Phi}}{\Phi}\right)_m
-\frac{w}{6\;m^2}\left(\frac{\dot{\Phi}}{\Phi}\right)_m^2\right]
,\label{pot}
\end{equation}
in which
\begin{equation}
\left(\frac{\dot{\Phi}}{\Phi}\right)_m=-2H_k\left(\frac{a_k}{a_m}\right)^3
\frac{1}{\sqrt{1+2w/3}}\frac{\Phi_k}{\Phi_m}\frac{\sqrt{\Phi_k-\Phi_m}}
{\sqrt{\Phi_k}},\label{ln}
\end{equation}
where the last expression is obtained by using Eq.(\ref{p}).

The value given by Eq.(\ref{pot}), coincides with that found in
general relativity theory, which is obtained by taking the limit
$\Phi=Cte.=\Phi_k=\Phi_m=m_p^2$ and $w\rightarrow \infty$
\cite{ureña}.

The ratio between the potential energies at the end of inflation
is given by

\begin{equation}
\frac{U_f}{V_f}=\frac{4\pi}{3}\frac{m^2\sigma_*^2}{\Phi_f\;
H_f^2}\ll \frac{m^2\;\Phi_m}{H_f^2\;\Phi_f }\left[1+\frac{1}{m}
\left(\frac{\dot{\Phi}}{\Phi}\right)_m
-\frac{w}{6\;m^2}\left(\frac{\dot{\Phi}}{\Phi}\right)_m^2\right]\label{u},
\end{equation}
and, in this way, the curvaton energy becomes subdominant at the
end of inflation or, equivalently
\begin{equation}
m^2\;\frac{\Phi_m}{\Phi_f}\left[1+\frac{1}{m}
\left(\frac{\dot{\Phi}}{\Phi}\right)_m
-\frac{w}{6\;m^2}\left(\frac{\dot{\Phi}}{\Phi}\right)_m^2\right]\ll
H_f^2\;. \label{one}
\end{equation}
In these expressions, we have used  $V_f=(3/8\pi) H_f^2\,\Phi_f$
and Eq.(\ref{pot}). Here, $\Phi_f$  is the  value  of the JBD
field at the end of inflation. Note that $U_f \ll V_f$ as it could
be seen from Eq.(\ref{u}).

At the time when the mass of the  curvaton field becomes
important, i.e. $m\simeq H$, its energy decays like a
non-relativistic matter in the form

\begin{equation}
\rho_\sigma=
\frac{m^2\sigma_*^2}{2}\frac{a_m^3}{a^3}\label{c_cae}.
\end{equation}

\subsection{Curvaton Decay After Domination\label{sectiv}}

As we have claimed, the curvaton decay could occur in two
different possible scenarios. In the first scenario, when the
curvaton comes to dominate the cosmic expansion (i.e.
$\rho_\sigma>\rho_\psi$), there must be a moment when the inflaton
and curvaton energy densities become equal. From Eqs.(\ref{rho}),
(\ref{h}) and (\ref{c_cae}) at the time when
$\rho_\sigma=\rho_\psi$, which happens when $a=a_{eq}$, we get

\begin{equation}
\left.\frac{\rho_\sigma}{\rho_\psi}\right|_{a=a_{eq}}=\frac{m^2\sigma_*^2}{2}\frac{a_m^3\;a_{eq}^3}{a_k^6\;\rho_\psi^k}
=\frac{4\pi}{3}\frac{m^2\sigma_*^2 a_m^3 a_{eq}^3}{H_k^2 \Phi_k
a_k^6} = 1,\label{equili}
\end{equation}
where we have used  $H_k^2=\frac{8\pi}{3\Phi_k}\rho_\psi^k$, since
we have that $\dot{\Phi}_k=\Phi'_k=0$.

Now from Eqs.(\ref{h}), (\ref{mh}) and (\ref{equili}), we may
write a relation for the Hubble parameter, $H(a_{eq})=H_{eq}$, in
terms of curvaton parameters,  the scale factor and the JBD scalar
field

\begin{eqnarray}
H_{eq}&=&
H_{k}\frac{a_k^3}{a_{eq}^3}\left[\frac{\sqrt{\Phi_k}}{\sqrt{1+2w/3}}
\;\frac{1}{\Phi_{eq}}\left(\sqrt{(\Phi_k-\Phi_{eq})}+
\sqrt{\Phi_k\left[1+\frac{2w}{3}\right]}\;\right)\right]\nonumber\\
&=&\frac{4\pi}{3}\frac{m\sigma_*^2}{(1+2w/3)}\frac{1}{\Phi_m
\Phi_{eq}}\left(\sqrt{(\Phi_k-\Phi_m)}+
 \sqrt{\Phi_k\left[1+\frac{2w}{3}\right]}\;\;\right)\left(\sqrt{(\Phi_k-\Phi_{eq})}+
 \sqrt{\Phi_k\left[1+\frac{2w}{3}\right]}\;\;\right).\label{heq}
\end{eqnarray}

This result should be compared to the corresponding result
associated with the general relativity theory, where
$H_{eq}^{(GR)}=4\pi\sigma_*^2 m/3m_p^2$.

On the one hand, the decay parameter $\Gamma_\sigma$ is
constrained by nucleosynthesis. For this, it is required that the
curvaton field decays before nucleosynthesis, which means
$H_{nucl}\sim 10^{-40}m_p < \Gamma_\sigma$. On the other hand, we
also require that the curvaton decay occurs after $\rho_\sigma >
\rho_\psi$, and $\Gamma_\sigma < H_{eq}$, so that we get a
constraint on the decay parameter, which is given by

\begin{equation}
10^{-40}m_{p}<\Gamma_{\sigma}< \frac{4\pi
m\sigma_*^2}{3(1+2w/3)}\frac{1}{\Phi_m
\Phi_{eq}}\left(\sqrt{(\Phi_k-\Phi_m)}+
 \sqrt{\Phi_k\left[1+\frac{2w}{3}\right]}\;\;\right)\left(\sqrt{(\Phi_k-\Phi_{eq})}+
 \sqrt{\Phi_k\left[1+\frac{2w}{3}\right]}\;\;\right),\label{gamm1}
\end{equation}
which, in the particular case when $\Phi_k\simeq\Phi_m$, we obtain
that $H_{eq}\simeq 4\pi\sigma_*^2 m
/3\Phi_{eq}\left(1+\frac{1}{\sqrt{1+2w/3}}\sqrt{\frac{\Phi_k-\Phi_{eq}}{\Phi_k}}\right)$.
Furthermore, if we demand that $w\gg1$ we find that $H_{eq}\simeq
4\pi\sigma_*^2 m /3\Phi_{eq}=H_{eq}^{(GR)}m_p^2/\Phi_{eq}$ and
therefore the decay parameter becomes constrained in the range
$10^{-40}m_p<\Gamma_\sigma<4\pi\sigma_*^2 m /3\Phi_{eq}$.

It is interesting to give an estimate of the constraint of the
parameters of our model, by using the scalar perturbation related
to the curvaton field. During the time where the fluctuations are
inside the horizon, they obey the same differential equation as
the inflaton fluctuations do, from which we conclude that they
acquire the amplitude $\delta\sigma_*\simeq H_*/2\pi$. On the
other hand, outside of the horizon, the fluctuations obey the same
differential equation as the unperturbed curvaton field and then,
we expect that they remain constant during inflation. The spectrum
of the Bardeen parameter $P_\zeta$, whose observed value is about
$2\times 10^{-9}$, allows us to determine the value of the
curvaton field $\sigma_*$ in terms of the parameters $\alpha$, $w$
and the JBD scalar field. At the time when the decays of the
curvaton fields occur, the Bardeen parameter becomes \cite{ref1u}

\begin{equation}
P_\zeta\simeq \frac{1}{9\pi^2}\frac{H_*^2}{\sigma_*^2}.
\label{pafter}
\end{equation}
The spectrum of fluctuations is automatically gaussian for
$\sigma_*^2\gg H_*^2/4\pi^2$, and is independent of
$\Gamma_\sigma$ \cite{ref1u}. This feature will simplify the
analysis in the space parameter of our model. Moreover, the
spectrum of fluctuations is the same as in the standard scenario.

>From expression (\ref{pafter}) and by using that
$H_*^2=H_f^2\Phi_fV_*/(\Phi_*V_f)$, we relate the perturbations to
the parameters of the model such that we get

\begin{equation}
V_0=27\pi^2(2-\beta)\frac{P_{\zeta}}{\alpha^2}\;\,\sigma_*^2\;e^{(\alpha\psi_*-N_*\beta)},
\label{18}
\end{equation}
where $N_*=\ln(a_f/a_*)$ is the number of the e-folds
corresponding to the cosmological scales, i.e. the number of
remaining inflationary e-folds at the time when the cosmological
scale exits the horizon. The last expression allows us to fix the
initial value of the effective potential  in terms of the free
parameters $\alpha$ and $w$. The constraint Eq. (\ref{one}) in
terms of $V_0$ becomes

\begin{equation}
m^2\left[1+\frac{1}{m} \left(\frac{\dot{\Phi}}{\Phi}\right)_m
-\frac{w}{6\;m^2}\left(\frac{\dot{\Phi}}{\Phi}\right)_m^2\right]\ll
\frac{8\pi}{3m_p^2\Phi_m}V_0 e^{(1-2/\beta)}.\label{21}
\end{equation}

\subsection{Curvaton Decay Before Domination\label{sectv}}

For the second scenario, the decay of the field happens before
this dominates the cosmological expansion. We need that the
curvaton field decays before its energy density becomes greater
than the inflaton one. Additionally, the mass of the curvaton is
non-negligible, so that we could use Eq. (\ref{c_cae}). The
curvaton decays at a time when $\Gamma_\sigma =H$ and then from
Eq. (\ref{h}) we get

\begin{equation}
\frac{\Gamma_\sigma}{H_k}=\frac{a_k^3}{a_d^3}\left[\frac{\sqrt{\Phi_k}}
{\sqrt{1+2w/3}}\;\frac{1}{\Phi_d}
\left(\sqrt{(\Phi_k-\Phi_d)}+
 \sqrt{\Phi_k\left[1+\frac{2w}{3}\right]}\;\right)\right], \label{Gamm}
\end{equation}
where ` d' labels the different quantities at the time when the
curvaton decays, allowing the curvaton field to decay after its
mass becomes important, so that $\Gamma_\sigma<m$; and before the
curvaton field dominates the expansion of the universe, i.e.,
$\Gamma_\sigma>H_{eq}$ (see Eq. (\ref{heq})).  Thus, we derive the
new constraints on the decay parameter, given by

\begin{equation}
\frac{4\pi}{3}\frac{m\sigma_*^2}{(1+2w/3)}\frac{1}{\Phi_m
\Phi_{eq}}\left(\sqrt{(\Phi_k-\Phi_m)}+
 \sqrt{\Phi_k\left[1+\frac{2w}{3}\right]}\;\;\right)\left(\sqrt{(\Phi_k-\Phi_{eq})}+
 \sqrt{\Phi_k\left[1+\frac{2w}{3}\right]}\;\;\right)<\Gamma_\sigma<m. \label{gamm2}
\end{equation}

Now, for the second scenario, the curvaton decays at the time when
$\rho_\sigma<\rho_\psi$. If we define the $r_d$ parameter as the
ratio between the curvaton and the inflaton energy densities,
evaluated at $a=a_d$ and for $r_d\ll 1$, the Bardeen parameter is
given by \cite{ref1u,L1L2}

\begin{equation}
P_\zeta\simeq \frac{r_d^2}{16\pi^2}\frac{H_*^2}{\sigma_*^2}.
\label{pbefore}
\end{equation}
With the help of Eqs. (\ref{mh}) and (\ref{Gamm}) we obtain

\begin{eqnarray}
r_d&=&\left.\frac{\rho_\sigma}{\rho_\psi}\right|_{a=a_d}=\frac{4\pi}{3}\frac{m^2\sigma_*^2
a_m^3 a_{d}^3}{H_k^2 \Phi_k
a_k^6} =\nonumber\\
&=&\frac{4\pi\sigma_*^2
}{3}\frac{m}{\Gamma_\sigma}\frac{1}{(1+2w/3)}\frac{1}{\Phi_m
\Phi_d}\left(\sqrt{(\Phi_k-\Phi_m)}+
 \sqrt{\Phi_k\left[1+\frac{2w}{3}\right]}\;\;\right)\left(\sqrt{(\Phi_k-\Phi_{d})}+
 \sqrt{\Phi_k\left[1+\frac{2w}{3}\right]}\;\;\right)\label{rd}.
\end{eqnarray}
Also, from Eq. (\ref{gamm2}) we get that

\begin{equation*}
 r_d<\frac{\Phi_{eq}}{\Phi_d}\frac{\left(\sqrt{(\Phi_k-\Phi_d)}+
 \sqrt{\Phi_k\left[1+\frac{2w}{3}\right]}\;\;\right)}{\left(\sqrt{(\Phi_k-\Phi_{eq})}+
 \sqrt{\Phi_k\left[1+\frac{2w}{3}\right]}\;\;\right)},
\end{equation*}
then, from $r_d\ll 1$, the last expression is written
\begin{equation*}
\frac{\Phi_{eq}}{\left(\sqrt{(\Phi_k-\Phi_{eq})}+
 \sqrt{\Phi_k\left[1+\frac{2w}{3}\right]}\;\;\right)}\ll\frac{\Phi_{d}}{\left(\sqrt{(\Phi_k-\Phi_{d})}+
 \sqrt{\Phi_k\left[1+\frac{2w}{3}\right]}\;\;\right)},
\end{equation*}

allowing us to use expression (\ref{pbefore}) for the Bardeen
parameter.

Expressions (\ref{pbefore}) and (\ref{rd}) could be written
\begin{equation}
\sigma_*^2=9\frac{P_\zeta}{m^2}
\frac{\Gamma_\sigma^2}{H_*^2}(1+2w/3)^2
\frac{\Phi_m^2}{\left(\sqrt{(\Phi_k-\Phi_{m})}+
 \sqrt{\Phi_k\left[1+\frac{2w}{3}\right]}\;\;\right)^2}\frac{\Phi_d^2}{\left(\sqrt{(\Phi_k-\Phi_{d})}+
 \sqrt{\Phi_k\left[1+\frac{2w}{3}\right]}\;\;\right)^2},
\end{equation}
and thus, expression (\ref{gamm2}) becomes

\begin{equation}
12\pi\frac{\Phi_m
\Phi_d^2}{\Phi_{eq}}\frac{\left(1+\frac{2w}{3}\right)}{\left(\sqrt{(\Phi_k-\Phi_{d})}+
 \sqrt{\Phi_k\left[1+\frac{2w}{3}\right]}\;\;\right)^2}
\frac{\left(\sqrt{(\Phi_k-\Phi_{eq})}+
 \sqrt{\Phi_k\left[1+\frac{2w}{3}\right]}\;\;\right)}{\left(\sqrt{(\Phi_k-\Phi_{m})}+
 \sqrt{\Phi_k\left[1+\frac{2w}{3}\right]}\;\;\right)}\;\frac{P_\zeta}{m^2
H_*^2}\;\Gamma_\sigma^2<\frac{\Gamma_\sigma}{m}<1.\label{newphys}
\end{equation}
The upper limit for the $\Gamma_\sigma$
 parameter, results to be

\begin{equation}
\Gamma_\sigma<\frac{m H_*^2\Phi_{eq}}{12\pi\Phi_m\Phi_d^2 P_\zeta}
 \frac{(\sqrt{\Phi_k-\Phi_{d}}+\sqrt{\Phi_k(1+2w/3)})^2
 (\sqrt{\Phi_k-\Phi_m}+\sqrt{\Phi_k(1+2w/3)})}{(1+2w/3)
 (\sqrt{\Phi_k-\Phi_{eq}}+\sqrt{\Phi_k(1+2w/3)})}\label{37}.
\end{equation}

In the limit $w\rightarrow \infty$ and
$\Phi_{eq}=\Phi_d=\Phi_m=\Phi_0=1/G=m_p^2$ the last expression
gives $\Gamma_\sigma<(12\pi P_\zeta)^{-1}mH_*^2G$, which
corresponds to the result obtained in the Einstein theory of
general relativity \cite{ureña}. Also, we see that
$\Gamma_\sigma<\left.\frac{m\Phi_k\Phi_{eq}H_*^2}{12\pi P_\zeta
\Phi_m\Phi_d^2}\right|_{w\gg 1}$. This tells us that, for $w\gg
1$, the upper limit for the $\Gamma_\sigma$ parameter depends on
the values of the JBD scalar field evaluated at the different
epochs.

\section{Constraints from gravitational waves\label{sectgw}}

The study of  gravitational waves that can be applied to our model
was described in Ref.\cite{Staro1}. It is interesting to give an
estimate of the constraint on the curvaton mass, using this type
of tensorial perturbation. Under the approximation give in
Ref.\cite{Staro2}, the corresponding gravitational wave amplitude
in the model by using JBD theory may be written as
\begin{equation*}
h_{GW}\simeq\,C_1\,H_*.
\end{equation*}
According to Ref. \cite{Dimo} we could have that $H\ll 10^{-5}m_p$
and $C_1$ is an arbitrary constant with unit of $1/m_p$. This
interesting thing comes from the fact that inflation could take
place at an energy scale smaller than grand unification. We note
this as an advantage of the curvaton approach against of the
single inflaton field scenario.

Now, using
$H^2_*=H_f^2\frac{\Phi_f}{\Phi_*}e^{-\alpha(\psi_*-\psi_f)}$, we
obtain
\begin{equation}
h_{GW}^2\simeq\,C_1^2
H_f^2\left(\frac{\Phi_f}{\Phi_*}\right)\;e^{-\alpha(\psi_*-\psi_f)}
  .\label{gw}
\end{equation}
In this way, from Eqs.(\ref{one}) and (\ref{gw}) we derive the
inequality
\begin{equation}
m^2\left[1+\frac{1}{m} \left(\frac{\dot{\Phi}}{\Phi}\right)_m
-\frac{w}{6\;m^2}\left(\frac{\dot{\Phi}}{\Phi}\right)_m^2\right]
\ll\,\frac{\Phi_*}{\Phi_m}\,\frac{h_{GW}^2}{C_1^2}\,
e^{\alpha(1/\alpha+\beta\alpha\Phi_*/8\pi)},\label{m2}
\end{equation}
and, in the particular case when $\Phi_k\simeq\Phi_m\simeq\Phi_*$,
from Eq. (\ref{ln}) we see that $(\dot{\Phi}/\Phi)_m\simeq 0$ and,
observing that Eq. (\ref{m2}) gives the following inequality: $m^2
\ll\,\frac{h_{GW}^2}{C_1^2}\, e^{\alpha(1/\alpha+\beta\alpha
m_p^2/8\pi)}$.


We note that we have $V_{,\psi\psi}=3\alpha^2 \Phi_f H_f^2/8\pi$
and, if according to Ref. \cite{Dimo}, if the inflaton field is
effectively massless, then its contribution to the curvature
perturbation will not be negligible and, in fact, could be in
excess related to the observational limit from COBE. This can be
avoided if the inflaton is massive, i.e. $V_{,\psi\psi}> H_f^2 $
or equivalently $\alpha^2>8\pi/3\Phi_f$, in which case its
perturbations are exponentially suppressed  as it could be seen
from this inequality.

In order to give an estimate of the gravitational wave, we move to
the kinetic epoch in which the energy density of gravitational
waves evolves as in Refs.\cite{Dimopoulus,referee4}:
\begin{equation}
\rho_g=\frac{32}{3\pi}h_{GW}^2\rho_\psi\left(\frac{a}{a_k}\right)^2\label{reata}.
\end{equation}

On the other hand,  when the curvaton field decays, i.e.
($\Gamma_\sigma \sim H$), it produces radiation which decays as
$1/a^4$. Then, when the curvaton decays, we may write for the
energy density
\begin{equation}
\rho_r^{(\sigma)}=\frac{m^2\sigma_*^2}{2}\frac{a_m^3}{a_d^3}
\frac{a_d^4}{a^4}.
\end{equation}

In order to keep the gravitational waves under control, we assume
that the radiation energy density is much larger than that
produced by inflation \cite{ureña}. Thus, at a time when
$a=a_{eq}$, we write

\begin{equation}
\left.\frac{\rho_r^{(\sigma)}}{\rho_\psi}\right|_{a=a_{eq}}=\frac{4\pi}{3\Phi_k}
\frac{m^2\sigma_*^2}{H_k^2}\frac{a_m^3}{a_k^3}
\frac{a_{eq}^2}{a_k^2}\frac{a_d}{a_k} =1.\label{qqq}
\end{equation}

Therefore, the constraint from the gravitational waves now reads
$$
\hspace{-12.0
cm}\left.\frac{\rho_g}{\rho^{(\sigma)}_r}\right|_{a=a_{eq}}=\frac{64}{3\pi}
h_{GW}^2\left(\frac{a_{eq}}{a_k}\right)^2
$$
\begin{equation}
\hspace{1.5cm}=\frac{16}{\pi^2} h_{GW}^2\frac{H_k}{m\sigma_*^2}
\left(\frac{\Gamma_\sigma}{H_k}\right)^{1/3}
\frac{\Phi_m\Phi_d^{1/3}\Phi_k^{1/3}}{[\sqrt{(\Phi_k-\Phi_m)}+\sqrt{\Phi_k(1+2w/3)}]}
\frac{(1+2w/3)^{2/3}}{[\sqrt{(\Phi_k-\Phi_d)}+\sqrt{\Phi_k(1+2w/3)}]^{1/3}}\ll
10^{-6},\label{ws}
\end{equation}
where we have used Eqs.(\ref{mh}), (\ref{Gamm}) and (\ref{qqq}).

We note that from Eq. (\ref{ws}) we obtain a bound for the mass of
the curvaton, $m$, given by

\begin{equation}
m\gg 10^7 h_{GW}^2 \frac{H_k}{\sigma_*^2}
\left(\frac{\Gamma_\sigma}{H_k}\right)^{1/3}
\frac{\Phi_m\Phi_d^{1/3}\Phi_k^{1/3}}{[\sqrt{(\Phi_k-\Phi_m)}+\sqrt{\Phi_k(1+2w/3)}]}
\frac{(1+2w/3)^{2/3}}{[\sqrt{(\Phi_k-\Phi_d)}+\sqrt{\Phi_k(1+2w/3)}]^{1/3}}
  \label{lachi}.
\end{equation}
We should note that, in this case, we have obtained a bound from
below for the corresponding curvaton mass, $m$.

\section{Conclusions \label{conclu}}

We have introduced the curvaton mechanism into a NO inflationary
model as another possible solution to the reheating problem in a
JBD theory.

In the context of the curvaton scenario, reheating does occur at
the time when the curvaton decays, but only in the period when the
curvaton dominates. In contrast, if the curvaton decays before its
density dominates the universe, reheating occurs when the
radiation due to the curvaton decay manages to dominate the
universe.

During the epoch in which the curvaton decay after that its
dominates ($\rho_\sigma >\rho_\psi$), the reheating temperature as
higher than $4.82\times 10^{-9}m_p$, since the decay parameter
$\Gamma_{\sigma}\propto\,T_{rh}^2/m_p$, where $T_{rh}$ represents
the reheating  temperature. Here, we have used Eqs. (\ref{gamm1})
and (\ref{pafter}), with $m\sim 10^{-8} m_p$, $H=10^{-8}m_p$,
$\Phi_m/\Phi_{eq}\sim\,1.25$, $\Phi_k/\Phi_{m}\sim\,1.33$,
$\Phi_{eq}=1.1 m_p^2$ and $w=3000$. For $w=500$ we obtain that the
temperature becomes $4.87\times 10^{-9}m_p$.

The result tells us that the reheating temperature becomes
practically independent with respect to the Jordan-Brans-Dicke
parameter $w$. Also the value that we have obtained $T_{rh}\leq
10^{-9}m_p$, agrees with the value obtained from gravitino
over-production, which gives $T_{rh}\leq 10^{-10}m_p$
\cite{rtref}.

In a JBD theory we have found that it is possible to use the
curvaton field for an effective exponential potential, i.e. for NO
models.  The dependence on the values of $w$ and the different
initial conditions for $\Phi_k$, $\Phi_m$ etc., permit us to reach
different values of the decay parameter $\Gamma_\sigma$, needed
for solving the problem of reheating in the NO models.

We can draw the allowed region for the parameter space, in a plot
of $m$ versus $\sigma_*$, for different conditions expressed by
the constraint on the model constraints (see Figure 1). Therefore
we plot only the constraint  Eqs. (\ref{pot}), (\ref{one}) and
(\ref{lachi}). The other constraints will be automatically
satisfied.  In this way, the curvaton mass becomes fine-tuned in
the sense that it depends on the values of the parameters used for
describing the corresponding cosmological models, as we could see
from expression (\ref{u}).

\begin{figure}[th]
\includegraphics[width=4.0in,angle=0,clip=true]{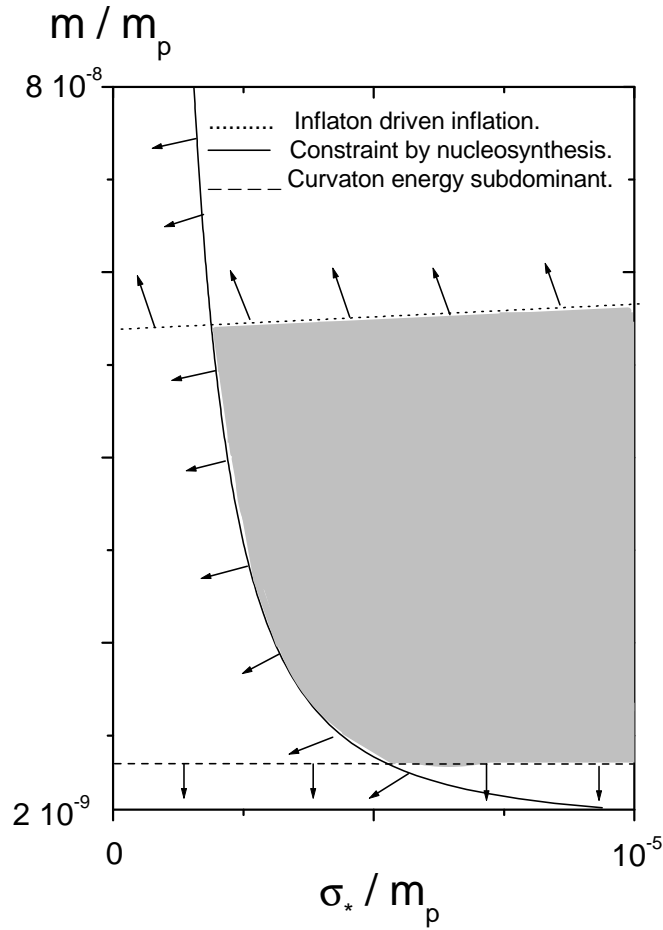}
\caption{Region of parameter space of model for $m$ versus
$\sigma_{*}$. The regions excluded by each constraint are
indicated by the arrows, and the allowed region is shaded. We have
taken the values : $w=3000$, $(\dot{\Phi}/{\Phi})_m=10^{-10}m_p$,
$\Phi_m/\Phi_{eq}=1.25$, $\Phi_{eq}=1.1 m_p^2$,
$V_0=10^{-11}m_p^4$ and $\Gamma_\sigma=10^{-18}m_p$. }
\end{figure}

The allowed region of the parameter space is reduced for smaller
values of the curvaton mass and  larger values of curvaton field.
This is in agreement with the fact that inflation could take place
at smaller - energy scales (smaller than the grand unification
scale).

\begin{acknowledgments}
CC was supported by MINISTERIO DE EDUCACION through MECESUP Grants
FSM 0204. SdC was supported by COMISION NACIONAL DE CIENCIAS Y
TECNOLOGIA through FONDECYT grants N$^0$ 1030469, N$^0$1040624,
N$^0$1051086 and also from UCV-DGIP N$^0$ 123.764, from
Direcci\'on de Investigaci\'on UFRO N$^0$ 120228 and from SEMILLA
1243.106. RH was supported by Grants PSD/06.

\end{acknowledgments}


\end{document}